\documentclass[a4paper,12pt,abstract=true,enabledeprecatedfontcommands]{scrartcl}

\usepackage{amsmath, amssymb}
\usepackage{tikz}
\usepackage[colorlinks=true,allcolors=blue]{hyperref}

\usepackage{authblk}

\begin{document}
\titlehead{\hfill KUNS-2695,~UCRHEP-T584}

\title{Ma-xion: Majoron as QCD axion\\ in a radiative seesaw model}

\author[1]{Ernest~Ma}
\affil[1]{Physics and Astronomy Department and Graduate Division, 
University of California, Riverside, California 92521, USA} 

\author[2]{Takahiro Ohata\thanks{Email: \texttt{tk.ohata@gauge.scphys.kyoto-u.ac.jp}}}
\affil[2]{Department of Physics, Kyoto University, Kyoto 606-8502, Japan}

\author[2]{Koji~Tsumura}
          
\maketitle

\begin{abstract}
The smallness of neutrino mass, the strong CP problem, and the existence of 
dark matter are explained in an economical way. 
The neutrino mass is generated by the colored version of 
a radiative seesaw mechanism by using color adjoint mediators. 
The Majorana mass term of the adjoint fermion, which carries lepton number $U(1)_{\mathbb{L}}$, 
is induced by its spontaneous breaking, resulting in a Majoron which doubles as the QCD 
(quantum chromodynamics) axion, thereby solving the strong CP problem. 
The breaking of $U(1)_{\mathbb{L}}$ sets simultaneously the seesaw scale for neutrino mass 
and the Peccei-Quinn breaking scale. 
This axion is a good candidate for dark matter as usually assumed. 
\end{abstract}

\section{Introduction}
%
The standard model of elementary particles (SM) has succeeded in describing 
high energy phenomena up to the TeV scale. 
However, there are several experimental and observational evidences of 
new physics beyond the SM, i.e. tiny neutrino masses, 
the existence of dark matter (DM) and dark energy, 
the density fluctuation from cosmic inflation, 
the baryon asymmetry of the universe and so on. 
From the theoretical point of view, 
variations of hierarchy problems, such as the strong CP problem, 
the naturalness of the Higgs boson mass, the cosmological constant, 
and the mass hierarchy of the SM fermions, are still issues to be 
explored and understood. 

As for the smallness of the neutrino mass, many seesaw mechanisms have been proposed. 
Since neutrinos may be Majorana particles\cite{Ref:Majorana}, many people focus on this possibility 
to explain the big differences between neutrino masses and the ordinary SM fermion masses. 
The Majorana neutrino mass is allowed by the unique dimension-five operator of the SM\cite{Ref:WeinbergOp}, 
which may be implemented by a new naturally large mass scale of the operator.
The simplest realization of this operator is the so-called type-I seesaw\cite{Ref:Type-I}, 
where the right-handed singlet partners of the SM neutrinos are introduced as mediators. 
Two more ways (type-II and type-III) exist to fulfil the seesaw mechanism at tree level\cite{Ref:Type-II,Ref:Type-III}. 
It was recognized many years ago\cite{Ref:TreeSeesaw} that these are the only three ways 
and the type-I,\,-II,\,and -III nomenclature was first introduced, together with the observation that 
there are generically also only three ways to realize this dimension-five operator 
in a one-particle-irreducible one-loop diagram.
Recently, a review (see \cite{Ref:vModelReview} and the references therein) of 
the many varieties of such radiative seesaw models appeared . 

There are many observational evidences of DM.  Its existence is no longer 
in doubt. 
On the other hand, no candidate particle is available within the SM. 
Whereas primordial black holes remain a possibility, this solution 
is likely to be ruled out by future observations and numerical studies\cite{Ref:PBHDM}. 
A new particle, sometimes introduced for a solution to a different problem of the SM, has been 
known as a good candidate for DM. 
Especially, a WIMP (Weakly Interacting Massive Particle) is a popular candidate, 
where its relic abundance is naturally fixed by thermal freeze-out\cite{Ref:WIMP}. 
The SIMP (Strongly Interacting Massive Particle) scenario is also 
spotlighted recently as the new candidate for thermal DM\cite{Ref:SIMP}. 
Many alternative candidates (WIMPzilla\cite{Ref:WIMPzilla}, Q-ball\cite{Ref:Qball}, and
axion\cite{Ref:AxionDM}) are also known in the broad range of the DM mass, 
where the right amount of DM can be achieved by nonthermal production. 

The only unobserved parameter in the SM is the QCD $\theta$ term. 
It is related to the chiral rotations of the quark fields through their mass terms, 
thus it is natural to expect a nonzero value. 
However, it has a tiny upper bound of $10^{-11}$\cite{Ref:theta-bound}, 
which is indicative of a fine-tuning problem or a new mechanism to forbid it. 
One way to solve the problem is to consider the massless up-quark\cite{Ref:MasslessMu}, 
where the $\theta$ term is rotated away. 
Another solution is the Peccei-Quinn(PQ) mechanism\cite{Ref:PQ}, 
where $\theta$ is promoted to a dynamical field\cite{Ref:WW}. 
The vanishing $\theta$ term is realized by its dynamical relaxation in 
the potential containing  
the vacuum expectation value (VEV) of the PQ symmetry breaking. 
%
 
In this Letter, we propose a new model which explains the smallness of neutrino mass, 
the strong CP problem and the existence of DM. 
The Majorana neutrino mass is generated by a one-loop radiative seesaw mechanism, 
where new color octet scalar and fermion fields circulate in the loop.
The lepton number conservation symmetry is identified as the PQ symmetry, 
and its spontaneous breaking produces a Majoron\cite{Ref:Majoron} 
as an axion for the  solution to the strong CP problem. 
This basic idea goes back many years~\cite{Ref:vPQ,Ref:vPQ2} and 
has recently been applied~\cite{mrz17} to a different axion model.
In this model, it gives rise to a Majorana mass term of the octet fermion. 
Therefore, the seesaw scale for neutrino mass and the PQ symmetry breaking are 
related to each other. 
The Majoron (QCD axion) is also used for DM as usual to make this a minimal model\cite{Ref:AxionDM}, 
although it is possible~\cite{dmt14} to have an additional WIMP candidate. 
%

This paper is organized as follows. 
In Section 2, the particle content and the brief sketch of our new model are given 
with the relevant Lagrangian terms. 
In Section 3, the neutrino mass generation mechanism, the solution to the strong CP problem, 
and the axion DM scenario are shown. 
The possible signatures and constraints of the model are also discussed. 
Conclusion and discussion are given in Section 4. 
%

\begin{table}[tb]
\centering
\begin{tabular}{|c||c|c|c|}
\hline
&$S$&$\Psi_{R}^{A}$&$\Phi_{}^{A}$\\
\hline\hline
$SU(3)_{C}$&$\bf{1}$&$\bf{8}$&$\bf{8}$\\
\hline
$SU(2)_{L}$&$\bf{1}$&$\bf{1}$&$\bf{2}$\\
\hline
$U(1)_{Y}$&$0$&$0$&$1/2$\\
\hline
$U(1)_{\mathbb{L}}$&$-2$&$1$&$0$\\
\hline
spin&$0$&$1/2$&$0$\\
\hline
\end{tabular}
\caption{New fields introduced to the Ma-xion model.}
\label{Tab:ParticleContent}
\end{table}

\section{Model}
%
To realize the PQ mechanism, colored fermions are needed which couple 
anomalously to $U(1)_{\text{PQ}}$, and the existence of a singlet scalar is also assumed 
which breaks it spontaneously.  In addition to the well-known  
KSVZ\cite{Ref:KSVZ} and DFSZ\cite{Ref:DFSZ} axion models, 
a third option exists in supersymmetry, using the gluino, i.e. a color octet fermion, 
assuming that its mass is dynamically generated\cite{Ref:GluinoAxion}. 
In this gluino axion model, the gluino plays the role of the heavy quark in the KSVZ model. 
The idea of our new model is to use a ``gluino'' for the neutrino mass generation. 

The particle content of the model is given in Table~\ref{Tab:ParticleContent}. 
A singlet scalar with the lepton number $\mathbb{L}=-2$ is added to 
the radiative seesaw model proposed by Fileviez Perez and Wise\cite{Ref:Fileviez Perez-Wise}, 
which is the color octet version of the simple scotogenic model\cite{Ref:Ma}. 
The color adjoint fermions $\Psi_{R}^{A}\,(A=1,2,\cdots, 8)$ and scalars $\Phi^{A}$ for 
the radiative seesaw mechanism are analogs to 
the right-handed neutrinos and the inert Higgs doublet 
in the scotogenic model.  Whereas an {\it ad hoc} dark parity was imposed originally 
to guarantee the stability of DM, it was shown more recently\cite{Ref:DarkParity,Ref:DarkParity2} 
that this dark parity is in fact derivable from lepton parity, 
a phenomenon applicable to many simple dark matter models proposed since 30 years ago. 
%
Unlike the scalar in the scotogenic model, the new colored scalar bosons may decay 
into the SM quarks through the Yukawa interactions:
\begin{align}
\mathcal{L}_{Q\Phi q_{R}^{}} 
=
g_{u}^{ij}\, \overline{Q_{i}}\, \widetilde{\Phi^{A}}\, T^{A}\, u_{jR}^{}
+g_{d}^{ij}\, \overline{Q_{i}}\, \Phi^{A}\, T^{A}\, d_{jR}^{}
+\text{H.c.}
\end{align}
where $\widetilde{\Phi^{A}}=i\,\sigma_{2} \Phi^{A\star}$, $i, j=1,2,3$ are the flavor indices, 
$g^{ij}_{q}\,(q=u, d)$ are the arbitrary Yukawa coupling constants, and 
the $SU(2)_{L}$ and $SU(3)_{C}$ indices are summed implicitly. 
For definiteness, we assume that $\Phi^{A}$ is much heavier than the weak scale, 
so that the flavor changing neutral current (FCNC) problem does not happen.\footnote{
If $g_{u,d}^{ij}$ are small enough while keeping the prompt decay of colored particles, 
$\Phi^A$ can become somewhat light. 
Further suppression of the FCNC is also possible by applying the minimal flavor violation hypothesis\cite{Ref:ManoharWise}, i.e. the Yukawa coupling matrices $g_{u,d}^{ij}$ are proportional to 
the quark Yukawa matrices $Y_{u,d}^{ij}=\sqrt2M_{u,d}^{ij}/v$ in the SM. } 
%
The new colored fermions also have Yukawa interactions with the SM left-handed lepton doublets 
and the scalar color octet:
\begin{align}
\mathcal{L}_{L\Phi\Psi_{R}} 
=
h_{\Psi}^{ij}
{\widetilde{\Phi}}^{A\dagger}
\overline{\Psi_{jR}^{A}}
L_i
+\text{H.c.}
\end{align}
where $h^{ij}_{\Psi}$ are the Yukawa coupling constants whose structure is related to 
the observed neutrino mass and mixing parameters\cite{Ref:vMass}. 
The lepton number of the colored fermions is determined through this interaction, i.e. $\mathbb{L}(\Psi_{R})=1$. 
In order to fit the observed neutrino oscillation data, at least two flavors of 
new Majorana fermions are required.
Hereafter, we assume three generations of gluino-like particles just for simplicity.  
The color octet $SU(2)_{L}$ doublet scalar field is parametrized as 
\begin{align}
\Phi^{A} = \begin{pmatrix} H^{+A} \\ (H^{A}+i\,A^{A})/\sqrt2 \end{pmatrix}. 
\end{align}
%
The Majorana mass term for the colored fermions is forbidden by the lepton number conservation, 
while the Yukawa interactions with the SM singlet scalar are allowed: 
\begin{align}
\mathcal{L}_{S\Psi_{R}\Psi_{R}} 
= 
-\frac12\, y_{\Psi}^{i}\, S\, \overline{(\Psi_{iR}^{A})^{c}} \Psi_{iR} + \text{H.c.}
\end{align}
Without any loss of generality, the Yukawa coupling matrix $y_{\Psi}^{}$ is taken to be diagonal. 
Indeed, the Majorana mass for each colored fermion is obtained after developing the VEV 
of the singlet, i.e., $M_{\Psi i} = y_{\Psi}^{i} \langle S \rangle$. 
Since the global lepton number symmetry is broken spontaneously, a Nambu-Goldstone boson, 
so-called Majoron, appears. 
Thanks to the existence of the new colored fermions (gluino-like particles), 
the Majoron is identified as an axion. 
Note that the lepton number symmetry $U(1)_{\mathbb{L}}$ plays the role of  
the $U(1)_{\text{PQ}}$ symmetry in this model. 
%

The model is a minimal setup to solve the strong CP problem, the existence of DM, and 
the smallness of neutrino masses at the same time. 
In the normal approach, the strong CP problem and the neutrino mass generation 
are considered as different problems, so that the mass scales are introduced separately 
for each problem. 
In our model, the seesaw scale and the PQ symmetry breaking scale have the common origin.\footnote{
In Ref.\cite{Ref:vPQ}(see also Ref.\cite{Ref:SMASH}), 
the identification of the PQ symmetry and the lepton number symmetry 
is discussed in the KSVZ realization with the type-I seesaw mechanism. 
In their model, the Majorana mass for right-handed neutrinos and the Dirac mass for 
the singlet heavy quark are arranged separately, but are generated by the VEV of the same singlet.  
}
The only energy scales introduced in our model are the negative mass squared of $S$ 
for the PQ symmetry breaking and the dimensionful parameter for $\Phi^{A\dag}\Phi^{A}$ term 
in addition to the one in the SM Higgs sector. 
%
From the viewpoint of the number of new fields, 
our model is comparable to the invisible axion models with the tree level seesaw mechanism. 
In addition to the common singlet field $S$ and a new mediator for the neutrino mass generation, 
singlet chiral quarks with different PQ charges are introduced in the KSVZ model, 
while two Higgs doublets are required in the DFSZ model. 
In all conventional models with the seesaw extension as well as in our model, 
three kinds of new particles are required.
%

The scalar potential of this model is given by
\begin{align}
\mathcal{V}
=&
-\mu^{2} H^{\dag}H -\mu_{S}^{2}S^{\star}S + M_{\Phi}^{2} \Phi^{A\dag} \Phi^{A} 
+\lambda (H^{\dag}H)^{2} \nonumber \\
&\quad
+\lambda_{S}^{} (S^{\star}S)^{2} 
+\lambda_{SH}^{} (S^{\star}S)(H^{\dag}H) 
+\lambda_{S\Phi} (S^{\star}S)\Phi^{A\dag} \Phi^{A} \nonumber \\
&\qquad
+\lambda_{3} (H^{\dag}H)\Phi^{A\dag} \Phi^{A} +\lambda_{4} |H^{\dag}\Phi|^{2}
+\frac12 \big\{ \lambda_{5} (H^{\dag}\Phi^{A})^{2} + \text{H.c.} \big\} 
+ \cdots 
\end{align}
where $H$ is the Higgs doublet in the SM.\footnote{
The complete scalar potential of $H$ and $\Phi^{A}$ can be found in Ref.\cite{Ref:ManoharWise}. 
}
As long as $M_{\Phi}^{2} \lesssim \lambda_{S\Phi}\langle S\rangle^{2}$, 
the mass of the new scalar doublet is controlled by the singlet VEV 
similarly to the DFSZ model. 
In this case, the model essentially has one new physics scale. 

\section{Solution to the Problems}

\begin{figure}
 \centering
 \includegraphics[width=0.5\textwidth]{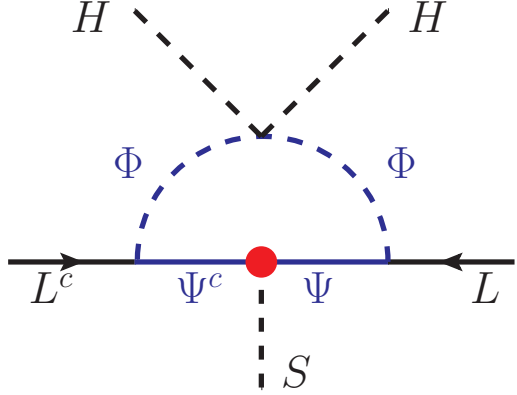}
 \caption{One-loop diagram for neutrino mass generation.}
 \label{Fig:ColoredMa}
\end{figure}

\noindent\underline{\bf{Neutrino mass}}\medskip\\
After developing the VEV of the singlet, the model arrives at the Perez-Wise model, 
where the neutrino mass is generated at the one-loop level with colored mediators. 
The Feynman diagram for the neutrino mass generation is given in Fig.\ref{Fig:ColoredMa}. 
By calculating this diagram, we obtain 
\begin{align}
(\mathcal{M}_\nu)_{ij}
=
- \frac{1}{4\pi^2} \sum_k h_{\Psi}^{ik}h_{\Psi}^{jk} M_{\Psi k} 
\Big( \frac{M_{H}^2}{M_{\Psi k}^2-M_{H}^2} \ln\frac{M_{H}^2}{M_{\Psi k}^2}
- \frac{M_{A}^2}{M_{\Psi k}^2-M_{A}^2} \ln\frac{M_{A}^2}{M_{\Psi k}^2} \Big), 
\end{align}
where the mass eigenvalues for the neutral component of the colored scalar are 
$M_{H,A}^{2}=M_{\Phi}^{2}+\frac12 \lambda_{S\Phi}^{}f_{a}^{2}+(\lambda_{3}+\lambda_{4} \pm \lambda_{5})v^{2}$.
The mass matrix takes the same form as the one in the scotogenic model\cite{Ref:Ma} 
up to the additional color factor of $8$. 
The structure of the mass matrix is easily maintained by the Yukawa coupling structure. 
The smallness of the neutrino mass is naturally explained not only 
by heavy colored particles but also by the radiative mechanism. 
Note that the magnitude of the Yukawa coupling constants $h_{\Psi}^{ij}$ and 
the mass squared difference of the colored scalars $(\propto \lambda_{5}^{})$ 
are additional sources of the suppression factor for the tiny neutrino mass.\footnote{
In a limit $2\lambda_5v^2\ll m_{0}^{2} =(M_{H}^2+M_{A}^2)/2$, the neutrino mass matrix is simplified as 
\begin{align}
(\mathcal{M}_\nu)_{ij}
&\simeq
\frac{1}{4\pi^2}\lambda_5v^2
\sum_k
h_\Psi^{ik}h_\Psi^{jk}
M_{\Psi k}
\frac{M_{\Psi k}^2\ln\frac{M_{\Psi k}^2}{m_0^2}-M_{\Psi k}^2+m_0^2}{(M_{\Psi k}^2-m_0^2)^2}.
\end{align}
For $2\lambda_5v^2\ll m_{0}^2 \ll M_{\Psi k}^2$ and
$\lambda_5\simeq1,\,h_{\Psi}^{ik}\simeq0.1,\,
y_\Psi^i\simeq1$,
the axion decay constant $f_a$ becomes $\mathcal{O}(10^{12})\,\mathrm{GeV}$.
The neutrino mass vanishes if we take one of three parameters, 
$\lambda_5, M_{\Psi k}, h_\Psi^{ik}$.  
Each of them corresponds to the symmetry violating parameter in the lepton number broken phase 
depending on the choices of the global $U(1)_{\mathbb{L}}$ charges of $\Psi_{R}^{A}$ and $\Phi^{A}$; 
\begin{align}
&\mathbb{L}(\Psi_R^A)=1,\;
\mathbb{L}(\Phi^A)=0&
&\Rightarrow
\quad 
\mathbb{L}(M_{\Psi k})\neq0,
\\
&\mathbb{L}'(\Psi_R^A)=0,\;
\mathbb{L}'(\Phi^A)=1&
&\Rightarrow
\quad 
\mathbb{L}(\lambda_5,\,g_{u,d}^{ij})\neq0,
\\
&\mathbb{L}''(\Psi_R^A)=0,\;
\mathbb{L}''(\Phi^A)=0&
&\Rightarrow
\quad 
\mathbb{L}(h_\Psi^{ik})\neq0.
\end{align}
Note that the operator $(H^\dagger\Phi^A)^2$ can be generated 
by the quark loop effect even if $\lambda_{5}$ is $0$ at tree-level. 
}
Utilizing this freedom to maintain the small neutrino mass, 
it is also possible to keep masses of the new colored particles in the TeV scale. 

Depending on how the neutrino mass is suppressed in the mass formula, 
varieties of the signature of the model are expected\cite{Ref:PhenoFW}. 
If the new colored particles are not super-heavy, 
the new colored particle production can happen at the high energy frontier machine. 
Especially, the same-sign dilepton signature (without missing energy) will  
probe the lepton number violating nature of the Majorana neutrino mass. 
The displaced-vertex signature due to the  long-lived color octet fermion 
is also interesting because it will probe the scale of the super-heavy mediator. 
At the luminosity frontier, searches for the charged lepton flavor violations 
$\ell_{i}\to\ell_{j}\gamma$ and the electroweak precision test are also useful 
to explore these heavy particles. 
These different searches obtain information on different parameters 
in the neutrino mass formula, and are thus complementary. \\
%

\noindent\underline{\bf{Strong CP problem}}\medskip\\
The effective axion-gluon-gluon coupling is generated by the triangle anomaly 
diagrams via the interaction between the Majoron and the color adjoint fermions, 
\begin{align}
{\mathcal L}_{a}
= 
-\frac{g^2}{32\pi^2} \Big(\theta-\frac{3\,n_{\Psi}^{}\,a(x)}{f_a}\Big) \tilde{G}^{A\mu\nu}G^A_{\mu\nu},
\end{align}
where we have also included the QCD $\theta$ term in the Lagrangian, and 
$n_{\Psi}^{}(=3)$ is the number of the color adjoint fermions. 
The gluon field strength tensor is $G^{A\mu\nu}$, $f_{a}$ is the axion decay constant, 
the axion field $a(x)$ is the phase of the electroweak singlet for the PQ symmetry breaking, i.e. 
$S(x)=\frac{1}{\sqrt{2}}\big(f_{a}+\sigma(x)\big)e^{i\,a(x)/f_{a}}$, 
and $\sigma(x)$ is a real scalar field with a mass of order $f_{a}$. 
A factor of $3$ in front of $n_{\Psi}^{}$ is the consequence of the adjoint representation.\footnote{
For one flavor of the fundamental representation, the factor is $1$ as in the KSVZ model. 
}
After the QCD phase transition, the axion potential becomes\cite{Ref:AxionPotential} 
\begin{align}
{\mathcal V_a}
= 
\Big(\frac{f_a}{3\,n_{\Psi}^{}}\Big)^2m_a^2 \Big\{1-\cos\Big(\theta-\frac{3\,n_{\Psi}^{}\,a(x)}{f_a}\Big)\Big\}, 
\label{Eq:AxionPotential}
\end{align}
by the non-perturbative effect of QCD. 
The axion mass is related to the decay constant similarly to the standard QCD axion as\cite{Ref:AxionMass}
\begin{align}
m_{a} \simeq 6\,\text{$\mu$eV} \times \Big(\frac{10^{12}\text{GeV}}{f_{a}/(3\,n_{\Psi}^{})}\Big).
\end{align}
By minimizing the axion potential, the CP invariance of the strong interaction is achieved dynamically. 
\\

\noindent\underline{\bf{Dark Matter}}\medskip\\
The axion is known as a candidate for cold DM. 
In the cosmic evolution, we assume that PQ symmetry breaking occurs before or during inflation. 
Under this assumption the axion field becomes homogeneous, 
so domain walls and axion strings are absent in our Universe. 
Thus the only process relevant to axion DM production is coherent oscillation due to the vacuum misalignment. 
The current axion energy density is given by\cite{Ref:AxionDM2,Ref:AxionDM3}
\begin{align}
\Omega_a h^2 \approx 
0.18\, \theta_{i}^{2}\, \Big( \frac{f_a/(3\,n_{\Psi}^{})}{10^{12}\text{GeV}} \Big)^{1.19}, 
\end{align}
where $h$ is the present Hubble parameter in units of $100\text{km/s/Mpc}$, and 
$\theta_{i}^{}$ is the initial axion misalignment angle, which takes the range $(-\pi,\pi)$. 
Since we assume that the PQ symmetry is broken before inflation ends, 
$\theta_{i}^{}$ takes the same constant value in the whole Universe and 
is considered as a free parameter. 
Hence the observed value $\Omega_\text{DM}h^2\sim 0.12$\cite{Ref:DMdensity} 
of the energy density for DM is easily explained. 
A robust lower bound on the decay constant $f_{a}/(3\,n_{\Psi}^{}) \gtrsim 4\times 10^{8}\,$GeV 
is known from the measured duration time of the neutrinos from the supernova SN 1987A\cite{Raffelt:2006cw}. \\
%

We note that the gluino axion model suffers from the cosmological domain wall problem\cite{Ref:DomainWall}, 
because the domain wall number is $N_{\text{DW}}=3\,n_{\Psi}^{}$ and cannot be one, as in the KSVZ model. 
If the inflation finishes before the PQ symmetry breaking, the axion field does not become 
homogeneous. As a result, domain walls are formed by the axion potential, Eq.\eqref{Eq:AxionPotential}. 
For this reason, it is necessary to assume that the PQ symmetry is broken before or during the inflation. 
Conversely, the color adjoint axion model can be verified if the inflation scale is determined by future observation. \\

A constraint can be derived from the isocurvature fluctuation. From Planck result\cite{Ref:DMdensity}, 
\begin{align}
\sqrt{
\mathcal{P}_S/
\mathcal{P}_\zeta
}
\lesssim
0.18, \qquad
\mathcal{P}_\zeta
\simeq
2.2\times10^{-9},
\end{align}
where $\mathcal{P}_S$ and $\mathcal{P}_\zeta$ are the dimensionless power spectrum of the DM isocurvature and curvature perturbations, respectively.
In our model, scalar $S$ has nonzero VEV during inflation, so that $\mathcal{P}_S$ becomes
\begin{align}
\mathcal{P}_S
\simeq
\bigg(
\frac{H_\mathrm{inf}}{
\pi(f_a/(3n_\Psi))\theta_i}
\bigg)^2
\bigg(
\frac{\Omega_ah^2}
{\Omega_\mathrm{CDM}h^2}
\bigg)^2,
\end{align}
where $H_\mathrm{inf}$ is the Hubble parameter during inflation. 
Therefore, $H_\mathrm{inf}$ is bounded to be 
\begin{align}
H_\mathrm{inf}
\lesssim
2\times10^7
\mathrm{GeV}\;
\theta_i^{-1}
\bigg(
\frac{10^{12}\text{GeV}}{f_a/(3\,n_{\Psi}^{})} \bigg)^{0.19}.
\end{align}

\section{Conclusion and Discussion}
We have constructed a model which explains the smallness of neutrino mass, 
the existence of cosmic DM, and the absence of strong CP violation at the same time. 
Color octet fermions (which carry lepton number) and scalars (which do not) are introduced to obtain Majorana 
neutrino masses by the radiative seesaw mechanism. 
In addition, a SM singlet scalar (which carries two units of lepton number) is chosen to break the lepton number symmetry dynamically. 
The color octet fermions obtain masses as a result, and the associated 
Goldstone boson plays the dual role of the Majoron as well as the QCD axion, 
because PQ symmetry is now identified with lepton number symmetry. 
The neutrino seesaw scale is thus also the PQ breaking scale. 
This axion is assumed to provide the necessary relic abundance to account for the DM of the Universe 
by a nonthermal production mechanism. 
This model also has the potential to explain other issues beyond the SM. 
The real component of the singlet scalar may be identified as the inflaton, 
whereas the decay of color octet fermions may be used to facilitate leptogenesis\cite{Ref:OctetLeptogenesis}. 
These topics are beyond the scope of this Letter, and are discussed elsewhere.

As an aside, we point out a possible realization of the PQ symmetry in the radiatively induced Dirac neutrino mass model~\cite{Ref:RadiativeDirac}.  
Leptoquark fields $\Phi_{\text{LQ}}$ and $\varphi$ are introduced to the KSVZ model 
so as to close the one-loop diagram for the neutrino mass generation. 
To be specific, the terms $\overline{L}(\Psi_{Q})_{R}\,i\sigma_{2}\Phi_{\text{LQ}}^{\star}$, 
$\overline{(\Psi_{Q})_{L}}N_{R}\varphi$, and $\Phi_{\text{LQ}}^{\dag}H\varphi$ are added, 
where $\Psi_{Q}$ is a color triplet vector-like fermion. 
By requiring Yukawa interactions (or the vanishing PQ charge) for $(\Psi_{Q})_{R}$ with SM particles, 
the PQ charges of $N_{R}, \Psi_{L}, S$ are determined to be the same and nonzero, 
which forbid the tree level neutrino mass automatically. 
An axion in this extension is no longer Majoron because of the lepton number conservation. 
Strong CP problem and the DM relic abundance and other topics beyond the SM 
can be explained in an analogous fashion. 
%

\section*{Acknowledgments}
The work of K. T. is supported by JSPS Grant-in-Aid for Young Scientists (B) (Grant No. 16K17697), 
by the MEXT Grant-in-Aid for Scientific Research on Innovation Areas (Grant No. 16H00868), 
and by Kyoto University: Supporting Program for Interaction-based Initiative Team Studies (SPIRITS).  
The work of E. M. is supported in part by the U. S. Department of Energy under Award No. DE-SC0008541. 

\end{document}